\documentstyle[twoside]{article}

\catcode`\@=11
\long\def\@makefntext#1{
\protect\noindent \hbox to 3.2pt {\hskip-.9pt
$^{{\eightrm\@thefnmark}}$\hfil}#1\hfill}               

\def\@makefnmark{\hbox to 0pt{$^{\@thefnmark}$\hss}}    

\def\ps@myheadings{\let\@mkboth\@gobbletwo
\def\@oddhead{\hbox{}
\rightmark\hfil\eightrm\thepage}
\def\@oddfoot{}\def\@evenhead{\eightrm\thepage\hfil
\leftmark\hbox{}}\def\@evenfoot{}
\def\sectionmark##1{}\def\subsectionmark##1{}}

\oddsidemargin=\evensidemargin
\newcounter{sectionc}\newcounter{subsectionc}\newcounter{subsubsectionc}
\renewcommand{\section}[1] {\vspace{12pt}\addtocounter{sectionc}{1}
\setcounter{subsectionc}{0}\setcounter{subsubsectionc}{0}\noindent
        {\tenbf\thesectionc. #1}\par\vspace{5pt}}
\renewcommand{\subsection}[1] {\vspace{12pt}\addtocounter{subsectionc}{1}
      \setcounter{subsubsectionc}{0}\noindent
      {\bf\thesectionc.\thesubsectionc.{\kern1pt \bfit #1}}\par\vspace{5pt}}
\renewcommand{\subsubsection}[1]
      {\vspace{12pt}\addtocounter{subsubsectionc}{1}
      \noindent{\tenrm\thesectionc.\thesubsectionc.\thesubsubsectionc.
      {\kern1pt \tenit #1}}\par\vspace{5pt}}
\newcommand{\nonumsection}[1] {\vspace{12pt}\noindent{\tenbf #1}
        \par\vspace{5pt}}

\newcounter{appendixc}
\newcounter{subappendixc}[appendixc]
\newcounter{subsubappendixc}[subappendixc]
\renewcommand{\thesubappendixc}{\Alph{appendixc}.\arabic{subappendixc}}
\renewcommand{\thesubsubappendixc}
        {\Alph{appendixc}.\arabic{subappendixc}.\arabic{subsubappendixc}}

\renewcommand{\appendix}[1] {\vspace{12pt}
        \refstepcounter{appendixc}
        \setcounter{figure}{0}
        \setcounter{table}{0}
        \setcounter{lemma}{0}
        \setcounter{theorem}{0}
        \setcounter{corollary}{0}
        \setcounter{definition}{0}
        \setcounter{equation}{0}
        \renewcommand{\thefigure}{\Alph{appendixc}.\arabic{figure}}
        \renewcommand{\thetable}{\Alph{appendixc}.\arabic{table}}
        \renewcommand{\theappendixc}{\Alph{appendixc}}
        \renewcommand{\thelemma}{\Alph{appendixc}.\arabic{lemma}}
        \renewcommand{\thetheorem}{\Alph{appendixc}.\arabic{theorem}}
        \renewcommand{\thedefinition}{\Alph{appendixc}.\arabic{definition}}
        \renewcommand{\thecorollary}{\Alph{appendixc}.\arabic{corollary}}
        \renewcommand{\theequation}{\Alph{appendixc}.\arabic{equation}}
        \noindent{\tenbf Appendix \theappendixc #1}\par\vspace{5pt}}
\newcommand{\subappendix}[1] {\vspace{12pt}
        \refstepcounter{subappendixc}
        \noindent{\bf Appendix \thesubappendixc. {\kern1pt \bfit #1}}
        \par\vspace{5pt}}
\newcommand{\subsubappendix}[1] {\vspace{12pt}
        \refstepcounter{subsubappendixc}
        \noindent{\rm Appendix \thesubsubappendixc. {\kern1pt \tenit #1}}
        \par\vspace{5pt}}

\newcommand{\smalllineskip}{\baselineskip=10pt}

\def\eightcirc{
\begin{picture}(0,0)
\put(4.4,1.8){\circle{6.5}}
\end{picture}}
\def\eightcopyright{\eightcirc\kern2.7pt\hbox{\eightrm c}}

\def\abstracts#1#2#3{{
        \centering{\begin{minipage}{4.5in}\baselineskip=10pt\footnotesize
        \parindent=0pt #1\par
        \parindent=15pt #2\par
        \parindent=15pt #3
        \end{minipage}}\par}}

\renewenvironment{thebibliography}[1]
        {\frenchspacing
         \ninerm\baselineskip=11pt
         \begin{list}{\arabic{enumi}.}
        {\usecounter{enumi}\setlength{\parsep}{0pt}
         \setlength{\leftmargin 12.7pt}{\rightmargin 0pt} 
         \setlength{\itemsep}{0pt} \settowidth
        {\labelwidth}{#1.}\sloppy}}{\end{list}}

\newcounter{itemlistc}
\newcounter{romanlistc}
\newcounter{alphlistc}
\newcounter{arabiclistc}

\newcommand{\fcaption}[1]{
        \refstepcounter{figure}
        \setbox\@tempboxa = \hbox{\footnotesize Fig.~\thefigure. #1}
        \ifdim \wd\@tempboxa > 5in
           {\begin{center}
        \parbox{5in}{\footnotesize\smalllineskip Fig.~\thefigure. #1}
            \end{center}}
        \else
             {\begin{center}
             {\footnotesize Fig.~\thefigure. #1}
              \end{center}}
        \fi}

\newcommand{\tcaption}[1]{
        \refstepcounter{table}
        \setbox\@tempboxa = \hbox{\footnotesize Table~\thetable. #1}
        \ifdim \wd\@tempboxa > 5in
           {\begin{center}
        \parbox{5in}{\footnotesize\smalllineskip Table~\thetable. #1}
            \end{center}}
        \else
             {\begin{center}
             {\footnotesize Table~\thetable. #1}
              \end{center}}
        \fi}

\def\@citex[#1]#2{\if@filesw\immediate\write\@auxout
        {\string\citation{#2}}\fi
\def\@citea{}\@cite{\@for\@citeb:=#2\do
        {\@citea\def\@citea{,}\@ifundefined
        {b@\@citeb}{{\bf ?}\@warning
        {Citation `\@citeb' on page \thepage \space undefined}}
        {\csname b@\@citeb\endcsname}}}{#1}}

\newif\if@cghi
\def\cite{\@cghitrue\@ifnextchar [{\@tempswatrue
        \@citex}{\@tempswafalse\@citex[]}}
\def\citelow{\@cghifalse\@ifnextchar [{\@tempswatrue
        \@citex}{\@tempswafalse\@citex[]}}
\def\@cite#1#2{{$\null^{#1}$\if@tempswa\typeout
        {IJCGA warning: optional citation argument
        ignored: `#2'} \fi}}

\def\@refcitex[#1]#2{\if@filesw\immediate\write\@auxout
        {\string\citation{#2}}\fi
\def\@citea{}\@refcite{\@for\@citeb:=#2\do
        {\@citea\def\@citea{, }\@ifundefined
        {b@\@citeb}{{\bf ?}\@warning
        {Citation `\@citeb' on page \thepage \space undefined}}
        \hbox{\csname b@\@citeb\endcsname}}}{#1}}

\def\@refcite#1#2{{#1\if@tempswa\typeout
        {IJCGA warning: optional citation argument
        ignored: `#2'} \fi}}

\def\refcite{\@ifnextchar[{\@tempswatrue
        \@refcitex}{\@tempswafalse\@refcitex[]}}

\def\pmb#1{\setbox0=\hbox{#1}
        \kern-.025em\copy0\kern-\wd0
        \kern.05em\copy0\kern-\wd0
        \kern-.025em\raise.0433em\box0}

\def\fnt#1#2{\footnotetext{\kern-.3em
        {$^{\mbox{\scriptsize #1}}$}{#2}}}

\def\fpage#1{\begingroup
\voffset=.3in
\thispagestyle{empty}\begin{table}[b]\centerline{\footnotesize #1}
        \end{table}\endgroup}

\def\runninghead#1#2{\pagestyle{myheadings}
\markboth{{\protect\footnotesize\it{\quad #1}}\hfill}
{\hfill{\protect\footnotesize\it{#2\quad}}}}
\font\tenrm=cmr10
\font\tenit=cmti10
\font\tenbf=cmbx10
\font\bfit=cmbxti10 at 10pt
\font\ninerm=cmr9

\font\eightrm=cmr8

\def\qed{\hbox{${\vcenter{\vbox{                      
   \hrule height 0.4pt\hbox{\vrule width 0.4pt height 6pt
   \kern5pt\vrule width 0.4pt}\hrule height 0.4pt}}}$}}

\begin{document}

\runninghead{D. Singleton}
{Asymptotic freedom of general relativity $\ldots$}

\thispagestyle{empty}\setcounter{page}{147}
\vspace*{0.88truein}
\fpage{147}

\centerline{\bf
ASYMPTOTIC FREEDOM OF GENERAL RELATIVITY AND}
\centerline{\bf ITS POSSIBLE
CONSEQUENCES}
\vspace*{0.035truein}

\vspace*{0.37truein}
\centerline{\footnotesize Douglas Singleton}

\centerline{\footnotesize \it
Department of Physics, CSU Fresno}
\baselineskip=10pt
\centerline{\footnotesize \it
2345 East San Ramon Ave., M/S 37}
\baselineskip=10pt
\centerline{\footnotesize \it
Fresno, CA 93740-8031 USA}
\baselineskip=10pt
\centerline{\footnotesize \it
E-mail: das3y@maxwell.phys.csufresno.edu}

\baselineskip 5mm

\vspace*{0.21truein}

\abstracts{The formation of singularities in certain situations,
such as the collapse of massive stars, is one
of the unresolved issues in classical general relativity.
Although no complete theory of quantum gravity
exists, it is often suggested that quantum gravity effects
may prevent the formation of these singularities. In this article
we will present arguments that a quantized theory of gravity
might exhibit asymptotic freedom.
Considering the similarites between non-Abelian gauge
theories and general relativity it is conjectured
that a quantized theory of gravity may have a coupling strength
which decreases with increasing energy scale. Such a scale dependent
coupling strength, could provide a concrete mechanism
for preventing the formation of singularities.}{}{}

\bigskip

$$$$

\section{\bf Introduction}

During the end stages in the evolution of certain supermassive stars
general relativity indicates that all
the material of the star will collapse into a singularity. This is
one of the difficulties with classical general relativity,
and it is often suggested that quantum gravity effects will somehow
prevent the formation of true singularities. Rhoades and Ruffini
\cite{ruff} have shown that even if the material of the star
``stiffens'' to the point where the speed of sound in the material
becomes equal to the speed of light, the formation of a singularity
can not be avoided if the star's final mass is $\ge
3.2 M_{\bigodot} \approx 6.4 \times 10^{30}$ kg.
The Hawking-Penrose theorems \cite{hp} show that
the formation of such singularities is a generic feature of classical
general relativity. A rough argument for why a singularity
inevitably forms for certain collapsing stars can be given
as follows : For a star in which the thermonuclear fire has gone
out, the gravitational attraction can be counterbalanced by the
quantum mechanical Pauli exclusion pressure. To get an
estimate of how the quantum mechanical pressure balances gravity
one can use energy considerations \cite{ohanian} with the
total energy of the star taken as the sum of the
energies of all the particles and the gravitational binding
energy. In the relativisitic regime
the average energy of each particle is on the order of
$c p_F$ where $p_F = (3 \pi ^2 \hbar ^3 N /V)^{1/3}$
is the Fermi momentum, and $N$ is the total number of particles
contained in the volume $V$. The total energy coming form this
source is $E_F = N c p_F$ or
\begin{equation}
\label{fermie}
E_F = N \left(3 \pi ^2 c^3 \hbar ^3 N  \over V \right) ^{1/3}
\end{equation}
The gravitational binding energy is of the order, $-G M^2 / R$,
where $M$ is the total mass and $R$ is the radius of the
star. Combining the gravitational binding energy and the
energy of Eq. (\ref{fermie}) gives an estimate for
the total energy of the system (ignoring the rest mass)
\begin{equation}
\label{energyt}
E_{total} = {\left(9 \pi c^3 N^4 \hbar ^3 \over 4 \right)} ^{1/3}
{1  \over R} - {G M^2 \over R}
\end{equation}
In the nonrelativistic case the first term goes as
$1/R^2$ \cite{ohanian} and there is a radius where stable equilibrium
is achieved. For the relativisitic case given in Eq.
(\ref{energyt}), the quantum mechanical exclusion pressure becomes
too ``soft''so that no stable equilibrium exists
and the star collapses. More rigorous work bears out the
conclusion of this rough estimate.
In addition to Ref. \cite{ruff}, Buchdahl
\cite{buch} \cite{wald} has shown that a
star of mass $M$ with a radius $R =9M / 4$ or smaller,
can not reach static equilibrium. These works indicate that
the formation of a singularity can not be prevented by the
mechanical forces that the material of the star could exert.

Another example of how general relativity results in particles
being inevitably forced to the central singularity of a gravitating
point mass can be seen by considering a test particle moving in the
Schwarzschild field of some point mass $M$. The effective potential
per unit mass is \cite{wald}
\begin{equation}
\label{effvm}
V_{eff} = {c^2 \over 2} - {G M \over r} + {L^2 \over 2 r^2} - {G M L^2 \over
c^2 r^3}
\end{equation}
The second term is the standard Newtonian gravitational potential per
unit mass, and the third term is the usual centripetal barrier. The last
term is a general relativistic addition. It has the effect that if
$r$ becomes too small there is no stable orbit, and the particle ends
up at $r=0$. This is to be contrasted with
Newtonian gravity where as long as $L^2 \ne 0$ the test particle will
not be pulled into the singularity

One option for avoiding these
singualrities is that gravity must somehow be
modified, and it is usually hypothesized that quantum gravity
will somehow accomplish this.
In particular one would like the strength of
the gravitational interaction to decrease at small distance,
or large energies. In the following sections we will
present various arguments that point to the possibility that
the gravitational interaction does decrease with decreasing
distance.

\section{\bf Scaling of $G$ by Analogy with Particle Physics}

Gauge theories play an important role in modern physics.
In the Standard Model \cite{SM} of particle physics,
matter particles interact via gauge
interactions of the group $SU(3) \times SU(2) \times U(1)$.
General relativity can also be cast in the form of a gauge
theory \cite{uti}. One key difference between the
gauge theories of particle physics and general relativity
is that the former have been successfully quantized, but
not the latter.

When the gauge theories of particle
physics are quantized certain phenomenon occur. In particular,
the coupling strength of the gauge theory becomes energy or
scale dependent. For the electromagnetic part of the Standard
Model the coupling strength increases with increasing energy
for the energies so far probed.
This is seen experimentally \cite{pdhb} where at low
energies $ e^2 / 4 \pi = \alpha _{em} \approx 1 / 137$ while at
energies around 100 GeV  $\alpha _{em} \approx 1/128$.
For the SU(3), strong interaction part of the Standard Model one finds
that the coupling strength decreases with increasing energy.
This decrease of the coupling strength with increasing  energy is
called asymptotic freedom \cite{politzer}, and its discovery
was one of the first successes of QCD, since it gave an explanation
for why, at high energies, the quarks inside the nucleon behaved
as if they were essentially free (Bjorken scaling) \cite{bjork}.
Thus, quantized gauge theories have coupling strengths
which are scale dependent, and non-Abelian gauge theories can exhibit
a coupling strength which becomes weaker at short distances or high
energies.

If general relativity is viewed as a gauge theory \cite{ramond}
one can speculate that, as in the case of other
quantized gauge theories, the coupling strength of a quantum
theory of gravity may become scale dependent. Since general
relativity shares similarities with non-Abelian gauge
theories, it could be conjectured that general relativity may
also be asymptotically free. One of the first theoretical
questions which any full theory of quantum gravity, such as
string theory or loop quantum gravity \cite{rovelli}, should
address is whether the gravitational interaction strength scales with
energy, and the nature of the scaling.

\section{\bf Scaling in effective theories of general relativity}

Effective field theory techniques allow one to
discuss the quantum corrections to field theories
even if the field theories are conventionally non-renormalizable.
Recently Donoghue \cite{don} applied effective field theory
methods to general relativity to calculate the lowest order
quantum corrections to the Newtonian potential.
For two point mass, $M_1$ and $M_2$ separated by a distance $r$,
quantum corrections modify the Newtonian potential as
\begin{equation}
\label{newtonv}
V(r) = - {G M_1 M_2 \over r} \left[ 1 - {127 \over 30 \pi ^2}{G \hbar
\over c^3 r^2} \right]
\end{equation}
In Ref. \cite{don} there is a further correction which goes as
$G(M_1 + M_2) / r c^2$. However this is just a post-Newtonian
correction from classical general relativity rather than a
quantum correction, since it does not contain $\hbar$. When a
correct theory of quantum gravity is found it should yield the
same kind of quantum corrections in the regime where the effective
field theory calculation is valid ({\it i.e.} for $r$ significantly
larger than the Planck length). From Eq. (\ref{newtonv}) it can be
seen that the quantum effects tend to decrease the strength of
the gravitational interaction as $r$ gets smaller. At ordinary
distances this effective decrease of the gravitational coupling
is currently unmeasurable since the second term in Eq. (\ref{newtonv})
is extremely small. It is possible to write Eq. (\ref{newtonv}) in the
usual form, $V(r) = - G (r) M_1 M_2 / r$, with a $r$ dependent $G$
\begin{equation}
\label{gr}
G(r) = G_{\infty} \left( 1 - {127 \over 30 \pi ^2} {G _{\infty} \hbar \over
c^3 r^2} \right)
\end{equation}
where $G_{\infty} \approx 6.67 \times 10 ^{-11} Nm^2/kg^2$
is the gravitational coupling constant determined as
$r \rightarrow \infty$. For distances not too close to the
Planck scale, Eq. (\ref{gr}) implies that Newton's
constant decreases with decreasing distance.
Usually the running of the coupling constant in gauge theories
is given in terms of a scaling with energy rather than with distance.
In the appropriate units one can replace distances $r$ for energies
$k$ via $r \rightarrow 1/k$, in terms of which the running $G$ from
Eq. (\ref{gr}) would become $G(k) = G_0 - (A G^2 _0) k^2$ \cite{reuter}
where $A > 0$ is a constant, and
$G_0 = G_{\infty}$ is the gravitational coupling
determined as $k \rightarrow 0$.

In terms of the effective potential of Eq. (\ref{effvm}) one can
replace $G$ by $G(r)$ of Eq. (\ref{gr}) so that the effective
potential becomes
\begin{equation}
\label{effvma}
{\tilde V}_{eff} = {c^2 \over 2} - {G (r) M \over r} + {L^2 \over 2 r^2} -
{G(r) M L^2 \over c^2 r^3}
\end{equation}
Now, whereas $V_{eff}$ of Eq. (\ref{effvm}) had no stable minimum
if $r$ became too small, ${\tilde V}_{eff}$ of Eq. (\ref{effvma}) always has
a stable minimum at some small $r$. This is most easily seen in the
$L = 0$ case where, by using Eq. (\ref{effvma}) to calculate
$d{\tilde V}_{eff} / dr = 0$, one finds that the effective potential
with the distance dependent $G$ has a minimum at
\begin{equation}
\label{rmin}
r_{min} = \sqrt{ 127 G_{\infty} \hbar \over 10 \pi ^2 c^3} \approx
1.8 \times 10 ^{-35} meters
\end{equation}
The numerical value for $r_{min}$ shows the weak point of
this hypothesized asymptotic freedom of general relativity :
this distance is at the Planck distance scale where the
effective theory used to calculate $G(r)$ of Eq. (\ref{gr})
is suspect. At this scale one really needs a full theory
of quantum gravity in order to calculate the scale
dependence of $G$ with confindence. One can still
speculate that this asymptotic freedom, indicated by the effective
field theory at low energies, continues to all energy scales for a
complete theory of quantum gravity. This is the reverse
of speculations in QCD, where the running of $\alpha _{QCD}$
in the high energy regime is often said to imply
the increase of the coupling strength at low energies, and therefore
confinement. Also it can be pointed out that superfically the effective
field theory result is not completely unreasonable. Plugging
$r _{min}$  back into Eq. (\ref{gr}) gives a value
for the second term of $\approx 0.3$ compared to the value of $1$ for the
first term, so that the first quantum correction is still smaller
than the zeroth order classical term.

To make a connection to the Fermi energy argument
we need to relate the distance $r$ between two point particles with the
radius $R$ of the star. For a star of radius $R$ composed of $N$
particles, the average distance between any two of the particles
will be roughly $r = R / (N)^{1/3}$. With this, Eq. (\ref{gr}) can
be written as
\begin{equation}
\label{gr1}
G(R) = G_{\infty} \left( 1 - {127 \over 30 \pi ^2}
{G_{\infty} \hbar N^{2/3} \over  c^3 R^2} \right)
\end{equation}
Replacing $G$ of Eq. (\ref{energyt}) with
the scale dependent $G(R)$ of Eq. (\ref{gr1})
and calculating $dE_{total} /dR$ now gives
\begin{equation}
\label{newequil}
{dE_{total} \over dR} = {1 \over R^2} \left( G_{\infty} M^2 - \left[
9 \pi c^3 \hbar^3 N^4 \over 4 \right] ^{1/3} \right)
-{127 G_{\infty} ^2 \hbar N^{2/3} M^2 \over 10 \pi ^2 c^3 R^4}
\end{equation}
The last term, which arises from the quantum corrections
of the effective gravitational field theory, ensures
that it is always possible to find some $R$ so that $dE/dR = 0$.
This implies that a stable balance between
gravity and the quantum mechanical pressure can be achieved due
to the weakening of the gravitational interaction.
Aside from the heuristic
nature of this argument (a more serious calculation would
use the Oppenheimer-Volkoff equation with the scale
dependent $G(R)$ of Eq. (\ref{gr1})) it is found
that the radius $R$, for which Eq. (\ref{newequil}) gives
an equilibrium, is again outside the regime where the effective field
theory calculation can be trusted. For a star
of mass $M = 5 M_{\bigodot} \approx 1.0 \times 10^{31} kg$
with $N = M/m_n = 5.97 \times 10 ^{57}$, it is
found that Eq. (\ref{newequil}) gives an equilibrium radius
of  $R = 2.70 \times 10 ^{-15}$ m. This implies an average spacing
between the particles of $r = R/N^{1/3} = 1.49 \times 10 ^{-34}$ m,
which is only one order of magnitude above the Planck scale of
$10^{-35}$ m. Again, one can hypothesize that the weakening of the
gravitational coupling, $G$, implied by the low energy effective
theory will continue at higher energy scales.

This hypothesized
asymptotic freedom for general relativity would not prevent the formation
of a black hole, since in the example given above the horizon forms at a distance
around 15 km. The scaling of $G$ only replaces the singularity at the
center of the black hole with an extemely dense, but non-singular mass.

\section{\bf Scaling in Kaluza-Klein Theories}

If the gravitational interaction is eventually unified
with the Standard Model interactions, then the scaling of the various
coupling strengths may be interrelated. This is similar
to grand unified theories such as SU(5) \cite{gg}, where the
scaling of the various non-gravitational
couplings are related. Kaluza-Klein theories offer
a simple and direct example of how the scaling of the gravitational
and non-gravitational coupling strengths may be related. In the
original Kaluza-Klein theory \cite{kal} a relationship
exists between the electric coupling $e$ and Newton's constant
\begin{equation}
\label{cc1}
G = {r_5 ^2 e^2 \over 16 \pi} = {r_5 ^2 \alpha _{em} \over 4}
\end{equation}
where $r_5$ was the radius of the curled up fifth dimension.
To get non-Abelian gauge fields it is
necessary to have more than one compactified dimension. In Ref.
\cite{weinberg} a relationship similar to Eq. (\ref{cc1}) is given
except with the electromagnetic coupling $e$ is replaced by the
non-Abelian coupling $g$, and $r_5$  replaced by some rms
circumference of the curled up dimensions. The key point about
Eq. (\ref{cc1}) or its non-Abelian version, is that Newton's constant
is proportional to the square of some non-gravitational coupling constant.
Thus $G$ should scale with distance or energy in the same manner
as $g^2$. For non-Abelian theories the coupling strength,
$\alpha  = g^2 / 4 \pi$, usually decreases in strength with
decreasing distance scale in a logarithmic way ({\it i.e.} $\alpha (r)
= \alpha _o [ 1 + c \alpha _o \ln (r / r_o)] ^{-1}$ where $c$ is
some positive constant which depends on the non-Abelian gauge group,
and $r_o$ is a reference distance at which the coupling is
measured) so that $G(r)$ should also decrease
logarithmically. The running of $G$ here is different
than in the previous section. First, as noted in Ref. \cite{reuter}
the running of $G$ implied by the effective field theory treatment
goes as a power of energy or inverse power of distance, whereas
in the present example, the running is logrithmic.
Secondly, in the effective field theory approach the direct quantum
corrections of gravity were
discussed. Here, the {\em direct} quantum effects of four dimensional
gravity are ignored, but one still finds that $G$ runs if  $g$
runs. At energies far from the Planck scale the description of the
compactified dimensions in terms of a non-Abelian
gauge field theory is reasonable, especially if these Kaluza-Klein
fields of the compactified dimensions are to describe real
non-Abelian fields. If the effective, non-Abelian
coupling, $g$, exhibits asymptotic freedom (as it should if it
is to model the behaviour of non-Abelian fields of the
Standard Model) then so will $G$. As in the previous section, when
Planck scale energies and distances are approached, this treatment of the
curled up dimensions by an effective non-Abelian gauge
theory breaks down, and a complete, non-perturbative method of
quantizing this higher dimensional gravitation theory is required.
This running of the gravitational coupling in Kaluza-Klein models
again opens up the possibility that the formation of singularities
in gravitational collapse may be avoided.

One worry about this Kaluza-Klein argument is that if
the running of the non-Abelian coupling, $g$, is
experimentally observed then the running of $G$ should also
be seen. For example, let $g$ be the QCD coupling. The
perturbative running of the QCD coupling is experimentally observed
at energy scales greater than about 2 GeV (see Ref. \cite{pdhb}
p. 82). If the running of $G$
were tied to $g$ then one might think that some experimental
signature of this running of $G$ should have be seen. In
accelerator experiments, however, one deals with such small quantities
of matter, gravitationally speaking, that any kind of running of
$G$ would be undetectable. Even inside an apparently high energy
environment like the interior of the Sun, where there is a gravitationally
significant amount of matter, one has a temperature
$\approx 1.6 \times 10 ^7$ K, which corresponds to an energy
scale of about 1.4 keV. This is not in the energy
range where the perturbative running of $g$ (and therefore
$G$) would apply.

Situations where a gravitationally significant amount of matter
at a high enough energy could exist, occur in situations of
gravitational collaspe. For example, taking a stellar mass of $M = 5
M_{\bigodot} = 1.0 \times 10 ^{31}$ kg, so that $N = M/m_n = 5.97
\times 10 ^ {57}$, and taking $R = 1000$ m gives, an average
energy per particle of $E_F / N \approx 6.9$ GeV
(see Eq. (\ref{fermie})), which is an energy
range where the perturbative scaling of $g$ should apply.

\section{\bf Conclusions}

We have argued that the singularities which occur in general
relativity in certain situations, such as gravitational
collapse, could possibly be avoided if quantum gravity exhibits
a scaling of Newton's constant. In the case of stellar collapse
it has been shown \cite{ruff} that even if the material of the
dead star exerts the maximum possible resistive force, it can not
counterbalance the inward push of gravity.
Thus the only obvious way to avoid these singularities would be to
somehow modify the gravitational interaction, which is essentially the
idea behind the common statements that a full quantum theory of gravity
would somehow prevent the formation of these singularities. In this
article we have presented motivations that a quantum theory of gravity
should have a coupling strength which weakens at short distance,
or large energy scales, thus allowing an equilibrium to be reached
between the quantum mechanical exclusion pressure and the weakened
gravitational interaction. First, by analogy with other
non-Abelian gauge theories, which when quantized exhibit asymptotic
freedom, we argued that a quantum theory of gravity may also exhibit
asymptotic freedom. Second, from recent effective field calculations
it is found that the gravitational interaction does grow weaker with
decreasing distance scale, at least
for scales which are not too close to the
Planck scale. Finally, if gravity is unified with the other interactions,
as in Kaluza-Klein theories, then the running of the different couplings
should be related; if the non-Abelian coupling $g$ exhibits asymptotic
freedom then so should $G$. All of these arguments are only good for
energies and distances far from the Planck scale. However, the idea
that quantum gravity may exhibit asymptotic freedom provides a concrete
mechanism of how the singularities of classical general relativity
may be avoided. There are currently theories, such as string theory or
loop quantum gravity, which hold out the hope of giving a
complete quantum theory of gravity. One of the first questions
that could be asked of such a complete theory of quantum gravity
would be the nature of the non-perturbative scaling, if any, that
it gives for $G$.

\section{\bf Acknowledgements} I would like to thank Vladimir
Dzhunushaliev for discussions and comments during the
writing of this article. This work was supported in part by
a Collaboration in Basic Science and Engineering grant from
the National Research Council.

\nonumsection{References}

\end{document}